\documentstyle[preprint,aps,floats,psfig]{revtex}
\catcode`@=11
\def\references{%
\ifpreprintsty
\bigskip\bigskip
\hbox to\hsize{\hss\large \refname\hss}%
\else
\vskip24pt
\hrule width\hsize\relax
\vskip 1.6cm
\fi
\list{\@biblabel{\arabic{enumiv}}}%
{\labelwidth\WidestRefLabelThusFar  \labelsep4pt %
\leftmargin\labelwidth %
\advance\leftmargin\labelsep %
\ifdim\baselinestretch pt>1 pt %
\parsep  4pt\relax %
\else %
\parsep  0pt\relax %
\fi
\itemsep\parsep %
\usecounter{enumiv}%
\let\p@enumiv\@empty
\def\theenumiv{\arabic{enumiv}}%
}%
\let\newblock\relax %
\sloppy\clubpenalty4000\widowpenalty4000
\sfcode`\.=1000\relax
\ifpreprintsty\else\small\fi
}
\catcode`@=12
\begin{document}
\def\mh{m_h^{}}
\def\vev#1{{\langle#1\rangle}}
\def\gev{{\rm GeV}}
\def\tev{{\rm TeV}}
\def\fbi{\rm fb^{-1}}
\def\lsim{\mathrel{\raise.3ex\hbox{$<$\kern-.75em\lower1ex\hbox{$\sim$}}}}
\def\gsim{\mathrel{\raise.3ex\hbox{$>$\kern-.75em\lower1ex\hbox{$\sim$}}}}
\newcommand{\hmu}{{\hat\mu}}
\newcommand{\hnu}{{\hat\nu}}
\newcommand{\hrho}{{\hat\rho}}
\newcommand{\hh}{{\hat{h}}}
\newcommand{\hg}{{\hat{g}}}
\newcommand{\hk}{{\hat\kappa}}
\newcommand{\tA}{{\widetilde{A}}}
\newcommand{\tP}{{\widetilde{P}}}
\newcommand{\tF}{{\widetilde{F}}}
\newcommand{\th}{{\widetilde{h}}}
\newcommand{\tp}{{\widetilde\phi}}
\newcommand{\tchi}{{\widetilde\chi}}
\newcommand{\te}{{\widetilde\eta}}
\newcommand{\vn}{{\vec{n}}}
\newcommand{\vm}{{\vec{m}}}

\newcommand{ \slashchar }[1]{\setbox0=\hbox{$#1$}   
   \dimen0=\wd0                                     
   \setbox1=\hbox{/} \dimen1=\wd1                   
   \ifdim\dimen0>\dimen1                            
      \rlap{\hbox to \dimen0{\hfil/\hfil}}          
      #1                                            
   \else                                            
      \rlap{\hbox to \dimen1{\hfil$#1$\hfil}}       
      /                                             
   \fi}                                             %

\tighten
\preprint{ \vbox{
\hbox{MADPH--00-1202}
\hbox{hep-ph/0011251}}}
\draft
\title{Resolving the Solar Neutrino Problem with KamLAND}
\author{V. Barger, Danny Marfatia and Benjamin P. Wood}
\vskip 0.3in
\address{Department of Physics, University of Wisconsin--Madison, WI 53706}
\vskip 0.1in
\maketitle

\begin{abstract}
{\rm We study how well KamLAND, the first terrestrial neutrino experiment capable of addressing the
 solar neutrino problem, will perform in ascertaining whether or not the large mixing angle
MSW solution (with \mbox{$10^{-5} \lsim \Delta m_{21}^2 \lsim 10^{-4}\, \rm{eV}^2$} and oscillation 
amplitude $\rm{sin}^2 2\,\theta_{12}>0.3$) 
is correct.  We find that in a year of operation KamLAND will provide unequivocal evidence for or against 
this solution.
 Furthermore, its sensitivity to the three-neutrino oscillation parameters in this region 
 is sufficiently acute as to determine $\Delta m_{21}^2$ to approximately $\pm 10$\% 
(for $\rm{sin}^2 2\,\theta_{12}>0.7$)  and to fix 
 $\rm{sin}^2 2\,\theta_{12}$ to within $\pm 0.1$ (at the $2\sigma$ level) with three years
of accumulated data, independent of the value of $\theta_{13}$.
}
\end{abstract}

\section{Introduction}

Several neutrino oscillation experiments now indicate that neutrinos are massive and that 
neutrino flavor mixing occurs. 
 Atmospheric neutrino experiments (Kamiokande~\cite{k1}, 
SuperKamiokande~\cite{sk1},
IMB~\cite{imb}, Soudan~\cite{soudan} and MACRO~\cite{macro}) 
report a $\nu_{\mu}/\nu_{e}$ event ratio that is about 
0.6 times the expected ratio; the $\nu_{\mu}$ flux shows 
a zenith angle dependence that is explained by oscillations.
All experiments that measure the solar neutrino
flux (Homestake~\cite{homestake}, SAGE~\cite{sage}, GALLEX~\cite{gallex}, Kamiokande~\cite{k} and 
SuperKamiokande~\cite{sk}) find a
deficit of $1/2$ to $1/3$ of the Standard Solar Model prediction \cite{SSM}.  LSND~\cite{lsnd},
an accelerator experiment, finds evidence for $\bar{\nu}_{\mu}\rightarrow \bar{\nu}_e$
and ${\nu}_{\mu}\rightarrow {\nu}_e$ oscillations, a result that is not excluded by KARMEN~\cite{karmen}
and awaits confirmation by MiniBooNE~\cite{miniboone}. 
The current generation of nuclear reactor experiments 
(Palo Verde~\cite{paloverde} and CHOOZ~\cite{chooz}) 
find null oscillation results
and rule out $\bar{\nu}_{e}\rightarrow \bar{\nu}_x$ oscillations for 
$\Delta m_{21}^2 \gsim 10^{-3}$ $\rm{eV}^2$ at maximal mixing and \mbox{$\rm{sin}^2 2\,\theta_{13} > 0.1$} 
for larger $\Delta m_{21}^2$ (at the 95\% confidence level). 

The focus of the present work is the solar neutrino puzzle \cite{bahcall} and KamLAND's~\cite{kamland} role in 
resolving whether or not the large mixing angle (LMA) solution (see {\it e.g.} Ref.~\cite{lma}) 
is the correct one. The LMA solution has $\Delta m_{21}^2$ in the range $10^{-5}$ to 
$10^{-4}$ $\rm{eV}^2$ 
and mixing amplitude  $\rm{sin}^2 2\,\theta_{12} > 0.3$. Data 
from SuperKamiokande \cite{osaka} now favors the LMA solution over the small mixing angle, low and
vacuum oscillation solutions (see Ref.~\cite{kras} for a discussion of the various oscillation solutions).
Solar neutrino measurements at SNO \cite{sno} should discriminate between different oscillation solutions,
but perhaps not in its first year of operation \cite{bahcall2}.
 If the LMA solution is the correct one, KamLAND will
provide a precise determination of the oscillation parameters.

KamLAND is unique in its potential as the first terrestrial experiment to explore
the solar neutrino anomaly. It will provide a definitive test of the LMA solution 
by either ruling it out or by pinning down the values of $\Delta m_{21}^2$ and $\rm{sin}^2 2\,\theta_{12}$.
KamLAND should be able to provide this answer in a year from the start of running in
spring 2001 and then provide an accurate determination of the solar neutrino oscillation parameters by the
end of the three years over which it is expected to take data for. Particularly significant
will be its ability to precisely determine $\Delta m_{21}^2$ in the LMA region 
 because unlike the case of solar
neutrinos, the $\Delta m_{21}^2$-dependent contribution to oscillations of the reactor neutrinos 
will not suffer from averaging of the oscillation $L/E$ dependence. 

 If the LMA solution is correct, $\Delta m_{21}^2$-dependent 
$CP$-violating effects \cite{CP0} 
can be large enough to be tested at very long baseline accelerator-based experiments~\cite{CP}. 
With KamLAND promising a precisely known value of  $\Delta m_{21}^2$, it will provide essential information
for future experiments studying  $CP$-violation in the neutrino sector.

In Section II we provide a brief overview of the KamLAND experiment and the oscillation hypothesis
in which we will work. Section III will be devoted to details of our simulation of the experiment
and the subsequent data analysis. We conclude in Section IV. 

\section{The KamLAND Experiment}

KamLAND is a reactor neutrino experiment with its
detector located at the Kamiokande site.  
With low energy neutrinos ($\vev{E_{\nu}} \sim 3$ MeV), it can only measure $\bar{\nu}_e$
disappearance and therefore will be unable to access small mixing
angles  $\rm{sin}^2 2\,\theta_{12} < 0.1$. About 95\% of the
$\bar{\nu}_e$ flux incident at KamLAND will be from reactors situated between
$80-350$ km from the detector, making the baseline long enough to provide a sensitive probe of 
the LMA solution of the solar neutrino problem. Specifically, the
sensitivity to the mass-squared differences 
will lie between $\Delta m_{21}^2 \sim (L\, ({\rm{m}})/ E\, (\rm{MeV}))^{-1} 
\gsim 10^{-5}$ ${\rm{eV}}^2$ and $\lsim 10^{-4}$ ${\rm{eV}}^2$. Despite the absence of a single baseline, 
the measured positron energy spectrum allows a sensitive probe of oscillation effects.

We consider a framework of three-flavor oscillations among the massive neutrinos $(\nu_1,\nu_2, \nu_3)$. 
In the phenomenologically interesting limit, $\Delta m^2\equiv m_2^2-m_1^2 \ll m_3^2-m_2^2$, the survival
probability at the detector is given by
\begin{equation}
P(\bar{\nu}_e\rightarrow \bar{\nu}_e)=1-2\,s_{13}^2\,c_{13}^2
- 4\,s_{12}^2\,c_{12}^2\,c_{13}^4\,{\rm{sin}^2} 
\bigg({1.27\, \Delta m^2 ({\rm{eV}^2})\, L ({\rm{m}}) \over E_{\nu} ({\rm{MeV}})}\bigg)\,,
\label{prob}
\end{equation}
where $s_{ij}^2=1-c_{ij}^2=\rm{sin}^2 \theta_{ij}$ are the neutrino mixing parameters, $L$ is the distance
of the detector from the source and $E_{\nu}$ is the energy of the anti-neutrino. 
Here we have averaged over the leading oscillations 
$\vev{{\rm{sin}^2} (1.27\, |m_3^2-m_2^2| \, L/ E_{\nu})}=1/2$.
Matter effects on oscillations are negligible at the KamLAND baseline. 

Although we restrict our analysis to three neutrino oscillations,
 it is possible that the transition $\bar{\nu}_e \rightarrow \bar{\nu}_{s}$, where $\nu_{s}$ is a sterile
neutrino, is responsible
for the depletion of the $\bar{\nu}_e$ flux. 
(See Refs.~ \cite{barger,sterile} for models and analyses  
 of four-neutrino oscillations for the solar and atmospheric anomalies). 
For example, in a $2+2$ neutrino mixing scheme, in which one pair
of nearly degenerate mass eigenstates has maximal ${\nu}_e \rightarrow \nu_{s}$ mixing for
solar neutrinos and the other pair has nearly maximal mixing ${\nu}_{\mu} \rightarrow \nu_{\tau}$
for atmospheric neutrinos,  the survival probability is \cite{barger}
\begin{equation}
P(\bar{\nu}_e\rightarrow \bar{\nu}_e)=1-4\,\epsilon^2
-{\rm{sin}^2} 
\bigg({1.27\, \Delta m^2 ({\rm{eV}^2})\, L ({\rm{m}}) \over E_{\nu} ({\rm{MeV}})}\bigg)\,,
\label{prob2}
\end{equation} 
where $\epsilon \simeq (0.016\,{\rm{eV}^2}/\Delta m_{LSND}^2)^{0.91}$ is restricted by 
 KARMEN~\cite{karmen} and BUGEY~\cite{bugey} to be in the interval $(0.01,0.1)$ 
corresponding to a $\Delta m_{LSND}^2$ range
 0.2 to 1.7 ${\rm{eV}^2}$ \cite{barger}. In Eq.~(\ref{prob2}) we have averaged over the 
leading oscillations associated with the LSND mass scale. 
On comparing Eq.~(\ref{prob}) (fixing $\theta_{12}=\pi/4$
and $\rm{sin}^2 2\,\theta_{13}=0.1$) with Eq.~(\ref{prob2}), one sees that KamLAND will be unable 
to distinguish whether the oscillations are to flavor or to sterile neutrinos, but neutral current
measurements at SNO will accomplish this.
KamLAND is more
 sensitive to $\theta_{13}$ than to $\epsilon$, and we find that it will not
provide useful information about $\theta_{13}$ and consequently even less about $\epsilon$. 
In the following, we base our analysis on Eq.~(\ref{prob}).

The  target for the $\bar{\nu}_e$ flux consists of a spherical transparent balloon filled with 1000
tons of non-doped liquid scintillator. The anti-neutrinos are detected via the inverse neutron $\beta$-decay
\begin{equation}
\bar{\nu}_e+p\rightarrow e^{+}+n\,.
\label{decay}
\end{equation} 
The cross section for this process is \cite{vogel}
\begin{eqnarray}
\sigma(E_{\nu})&=&{2\,\pi^2 \over m_e^5\,f\,\tau_n}(E_{\nu}-\Delta M)\,[(E_{\nu}-\Delta M)^2-m_e^2] 
\nonumber\\&=&
0.952\,{(E_{\nu}-\Delta M)\,[(E_{\nu}-\Delta M)^2-m_e^2] \over 1\,\rm{MeV}^2}\times 10^{-43}\, \rm{cm}^2\,,
\label{cross}
\end{eqnarray} 
where $m_e$ is the mass of the electron, $\Delta M=m_n-m_p$ is the neutron-proton mass difference,
$\tau_n=(886.7 \pm 1.9)$ s is the neutron lifetime and 
$f=1.7152$ is the phase space factor which includes Coulomb, weak magnetism, recoil and outer 
radiative corrections. The theoretical error in the above cross section is less than a percent. The neutrino
capture process has a threshold of \mbox{$\Delta M+ m_e = 1.804$ MeV} 
and the $\bar{\nu}_e$ flux above this threshold
is $1.3 \times 10^6$ $\rm{cm}^{-2}\,\rm{s}^{-1}$ which is known to a precision of about 1.4\%
\cite{french}.
In the case of no oscillations, KamLAND expects to see 
$\sim$ 800 events per year with a background $\sim$ 40 events per year. The distribution of the background
events versus positron energy is expected to be known, thus facilitating a clean extraction 
of the signal. We will assume that a background subtraction can be made 
for our data simulation and analysis.

\section{Data Simulation and Analysis}

To calculate the $\bar{\nu}_e$ flux we include contributions from all nuclear reactors within a 
radius of 350 km from the
detector.  Table \ref{locations} gives the relative fluxes (without oscillations) 
and thus shows the relative importance of each reactor to the experiment (total fluxes 
from each reactor are tabulated in Ref.~\cite{kamland}). 
The $\bar{\nu}_e$ spectrum above the 1.8 MeV threshold for the process (\ref{decay}) 
is the result of the decay of fission fragments of the isotopes $^{235}\rm{U}$, $^{239}\rm{Pu}$, 
$^{238}\rm{U}$ and  $^{241}\rm{Pu}$. The spectrum from the fission products of  
 each of these isotopes can be found in \cite{nuc}. As the reactor operates, 
the concentration of  $^{235}\rm{U}$
decreases and that of $^{239}\rm{Pu}$ and $^{241}\rm{Pu}$ increases. We do not account for the
fissile isotope evolution and instead assume a typical fraction of fissions (as in Ref.~\cite{french}) 
from the four fissile materials (see Table \ref{fission}). 
We have confirmed that the effect of the evolution is small by reproducing the $\bar{\nu}_e$
spectrum from the Palo Verde experiment \cite{paloverde}.
\begin{table}[t]
\begin{eqnarray}
\begin{array}{lcccr}
\rm{Reactor}  &  &  \rm{Distance}\, (\rm{km}) &  & \rm{Percent\, of\, total\,flux}  \\
\hline
\rm{Kashiwazaki}   &  &  160.0                 &  & 33.36 \\
\rm{Ohi}           &  &  179.5                 &  & 14.76  \\
\rm{Takahama}      &  &  190.6                 &  & 9.76  \\
\rm{Shiga}         &  &   80.6                 &  & 8.51  \\
\rm{Tsuruga}       &  &  138.6                 &  & 8.12  \\
\rm{Mihama}        &  &  145.4                 &  & 8.10  \\  
\rm{Hamaoka}       &  &  214.0                 &  & 8.06  \\
\rm{Fukushima-1}   &  &  344.0                 &  & 4.17  \\
\rm{Fukushima-2}   &  &  344.0                 &  & 3.86  \\
\rm{Tokai-II}      &  &  294.6                 &  & 1.30 \nonumber
\end{array}
\end{eqnarray}
\caption[]{The expected relative contribution of each reactor to the $\bar{\nu}_e$ flux detected at
KamLAND in the case of no oscillations. Only reactors in a 350 km radius of the detector
are considered.   }
\label{locations}
\end{table}
\begin{table}[t]
\begin{eqnarray}
\begin{array}{lcccccc}
                   &  &  \rm{Start\, of\, Cycle}\,(\%)
&  & \rm{End\, of\, Cycle}\,(\%) 
& &  \rm{Typical\, value}\,(\%) 
 \\
\hline
^{235}\rm{U}       &  &  60.5                 &  & 45.0  & &  53.8 \\
^{239}\rm{Pu}      &  &  27.2                 &  & 38.8  & &  32.8 \\
^{238}\rm{U}       &  &  7.7                  &  & 8.3   & &  7.8  \\
^{241}\rm{Pu}      &  &  4.6                  &  & 7.9   & &  5.6 \nonumber
\end{array}
\end{eqnarray}
\caption[]{The fraction of fissions from the four fissile elements in a nuclear reactor at the
beginning of the cycle, the end of the cycle and a typical value during a cycle taken from Ref.~\cite{french}.  }
\label{fission}
\end{table}

With the relative flux from each reactor known, and the knowledge of the  $\bar{\nu}_e$
spectrum (assumed to be the same for all reactors), we calculated the 
$e^+$ spectrum at KamLAND resulting from inverse $\beta$-decay and normalized it to yield a total of
800 events per year in the absence of oscillations. This procedure effectively accounts for 
the effective number of free protons in the target, the $e^+$ and $n$ efficiencies, the
efficiency of the $e^+ - n$ distance cut and fluctuations in the power output of the reactors arising
 from dead time for maintenance and seasonal variations of power requirements.

\begin{figure}[t]
\mbox{\psfig{file=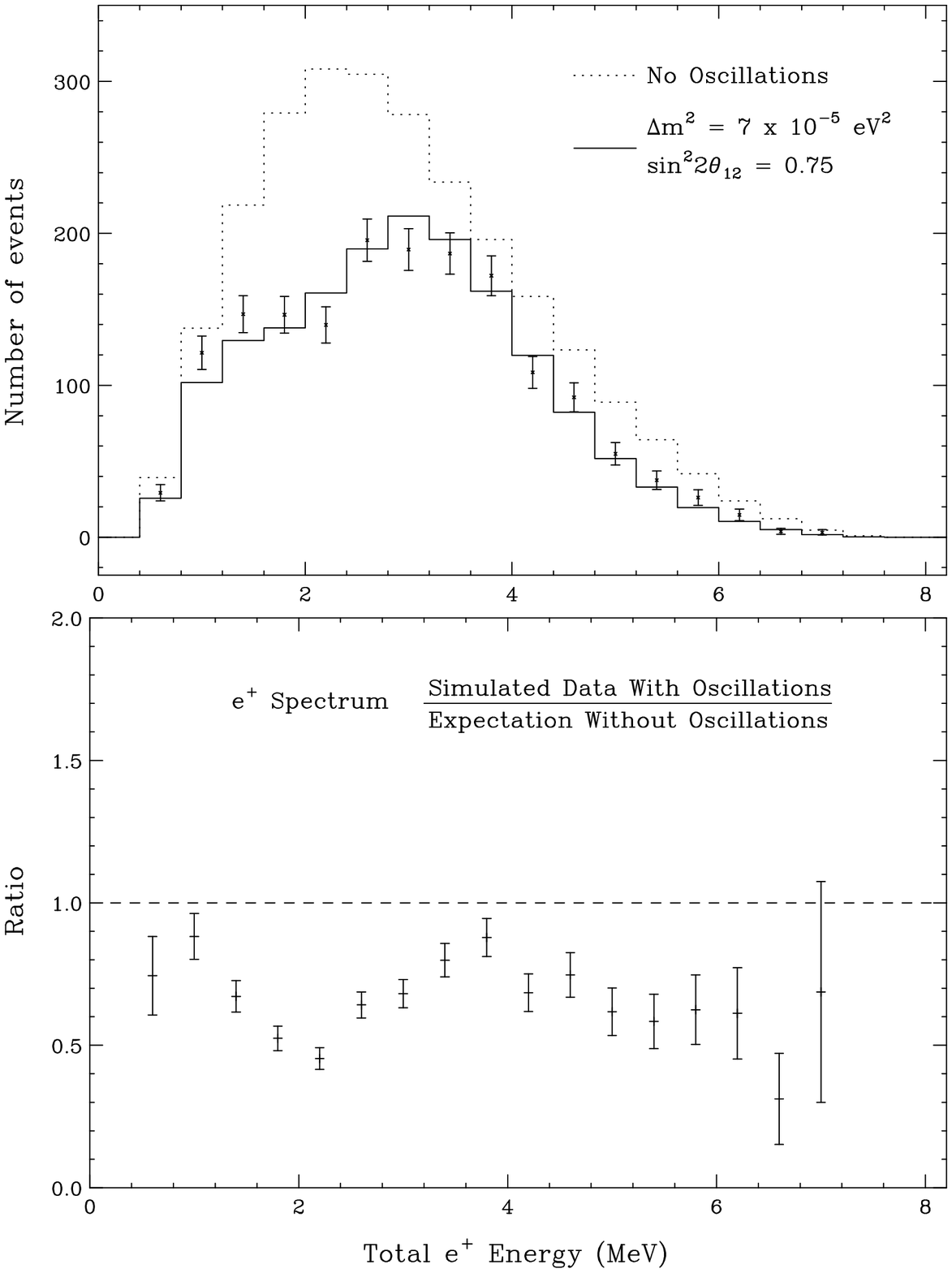,width=15cm,height=15cm}}
\caption{The expected $e^+$ energy spectrum (versus total $e^+$ energy) with three years of accumulated 
data for the case of no oscillations (dotted histogram) and for
\mbox{$(\Delta m^2, \rm{sin}^2 2\,\theta_{12})$=($7 \times 10^{-5}\, \rm{eV}^2, 0.75$)}
(solid histogram). The data points represent the simulated
spectrum.  The lower plot shows the
ratio of the simulated data to the expectation for no oscillations. 
The errors shown are statistical.}
\label{kamexp}
\end{figure}

\begin{figure}[tb]
\mbox{\psfig{file=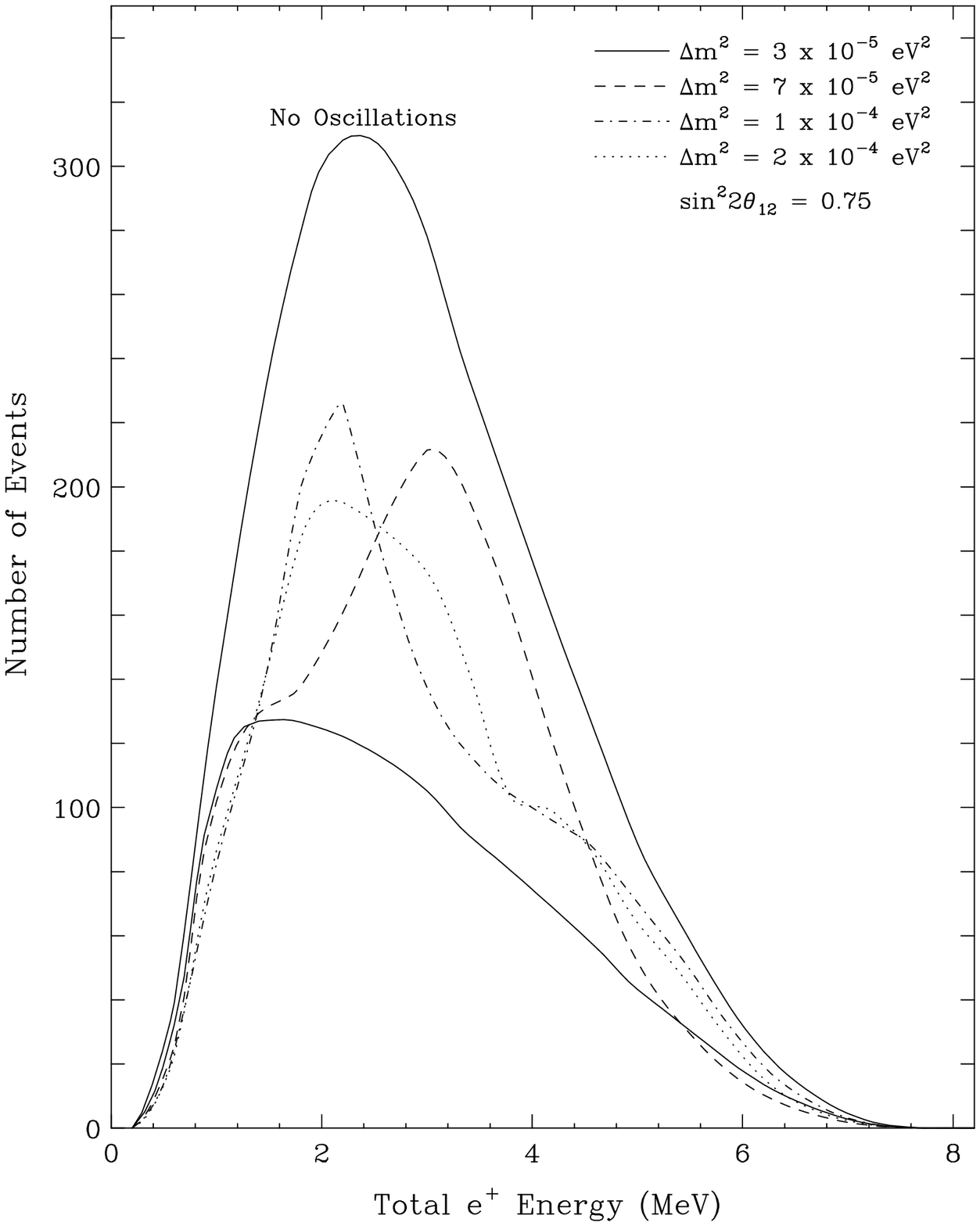,width=15cm,height=15cm}}
\caption{KamLAND's sensitivity to $\Delta m^2$ is unprecedented. In three years 
it will easily be able to discriminate between only slightly different 
values of $\Delta m^2$ in the LMA region. For $\Delta m^2 \gsim 2 \times 10^{-4}\, \rm{eV}^2$ the spectra
overlap significantly.}
\label{kamosc}
\end{figure}

Figure~\ref{kamexp} shows the $e^+$ energy spectrum expected at KamLAND with three years of data,  
illustrated for 
the cases of no oscillations (dotted histogram) and 
\mbox{$(\Delta m^2, \rm{sin}^2 2\,\theta_{12})$=($7 \times 10^{-5}\, \rm{eV}^2, 0.75$)}
(solid histogram). Each simulated data point is generated by randomly choosing a point from a Gaussian 
distribution centered at the theoretical
value and of width equal to the square root of the theoretical value.  The lower plot shows the
ratio of the simulated data to the expectation for no oscillations. 
The errors shown are statistical. The plots are versus the total $e^+$ energy.
A plot of the total measurable energy would be shifted to the right by $m_e=0.511$ MeV
 because the $\bar{\nu}_e$ signature involves a measurement of the kinetic
energy of the $e^+$ and the annihilation energy in the form of two 0.511 MeV gamma rays.
Figure~\ref{kamosc} illustrates the significant changes in the  $e^+$ spectrum for 
 different values of $\Delta m^2$
at a given value of  $\rm{sin}^2 2\,\theta_{12}$. The spectra overlap to a large extent 
for values of $\Delta m^2 \gsim 2 \times 10^{-4}\, \rm{eV}^2$.

For the statistical analysis we define $\chi^2$ as
\begin{equation}
\chi^2(\Delta m^2,{\rm{sin}^2} 2\,\theta_{12}) = \sum_{i=1}^{17} 
{(N_{{\rm{simulated}}}(E_i)-N(E_i,\Delta m^2,{\rm{sin}^2} 2\,\theta_{12}))^2 \over N_{{\rm{simulated}}}(E_i)} \,,
\end{equation}
where $i$ labels the 17 $e^+$ energy bins each of width $0.4$ MeV and midpoint $E_i$, 
 $N_{\rm{simulated}}(E_i)$ is the number of simulated events in the $i^{\rm{th}}$ bin
 and $N(E_i,\Delta m^2,\rm{sin}^2 2\,\theta_{12})$ is the corresponding 
theoretical value for oscillation parameters $(\Delta m^2, \rm{sin}^2 2\,\theta_{12})$.
 The only fitted parameters are  $\Delta m^2$ and  $\rm{sin}^2 2\,\theta_{12}$. We keep
 $\rm{sin}^2 2\,\theta_{13}$ fixed when performing the  $\chi^2$ analysis (using the
Levenberg-Marquadt method). From CHOOZ we know that  $\rm{sin}^2 2\,\theta_{13}<0.1$ and
we consider the two extreme cases $\rm{sin}^2 2\,\theta_{13}=0$ and 
$\rm{sin}^2 2\,\theta_{13}=0.1$. 

It is important to note that the main source of uncertainty
in the experiment comes from conversion of the fission rates in the reactors to  $\bar{\nu}_e$
fluxes and the normalization uncertainty is expected to be less than 3\%
\cite{shirai}. (The CHOOZ experiment~\cite{chooz} had a normalization uncertainty of 2.7\%). 
Even if the combined systematic uncertainty could be as large as  5\%,
 the shapes and sizes of the
confidence contours in our analysis would be unaffected by its inclusion. 
We can thus safely ignore systematic uncertainties in what follows. 

We concentrate on the region defined by  $\rm{sin}^2 2\,\theta_{12}>0.2$ and
\mbox{$10^{-5}<\Delta m^2<2 \times 10^{-4}$ $\rm{eV}^2$.} 
For values of $\Delta m^2$ outside this region, sensitivity to the neutrino energy-dependence 
is lost. 
For \mbox{$\Delta m^2 \lsim 10^{-5}$  $\rm{eV}^2$}, the value of the 
oscillation-dependent sinusoidal factor gets small.
For $\Delta m^2 \gsim 2 \times 10^{-4}$ $\rm{eV}^2$, 
the argument of the $\Delta m^2$-dependent sine function in Eq.~(\ref{prob}) becomes large because of the 
long baseline and the
oscillations get averaged. By studying the behavior of $\chi^2-\chi^2_{min}$ as a function $\Delta m^2$
for several simulated datasets, we find that for theoretical 
inputs with $\Delta m^2_{theor} \lsim 2 \times 10^{-4}\, \rm{eV}^2$,
$\chi^2-\chi^2_{min}$ has unique, well-defined minima at the theoretical values.
 For inputs with $\Delta m^2_{theor} \gsim 2 \times 10^{-4}\, \rm{eV}^2$ 
we find the presence of more than one acceptable minimum in the region 
$2 \times 10^{-4} \lsim \Delta m^2 \lsim 10^{-3}\, \rm{eV}^2$ \cite{barb} and a continuum of solutions 
\mbox{$\Delta m^2 \gsim 10^{-3}\, \rm{eV}^2$}
that are acceptable at the $2\sigma$ level. Thus, for values larger than
$2 \times 10^{-4}\, \rm{eV}^2$, KamLAND can place a lower bound on $\Delta m^2$, but
cannot uniquely determine the value.
However, the region of good sensitivity covers the entire LMA solution.

\begin{figure}[tb]
\mbox{\psfig{file=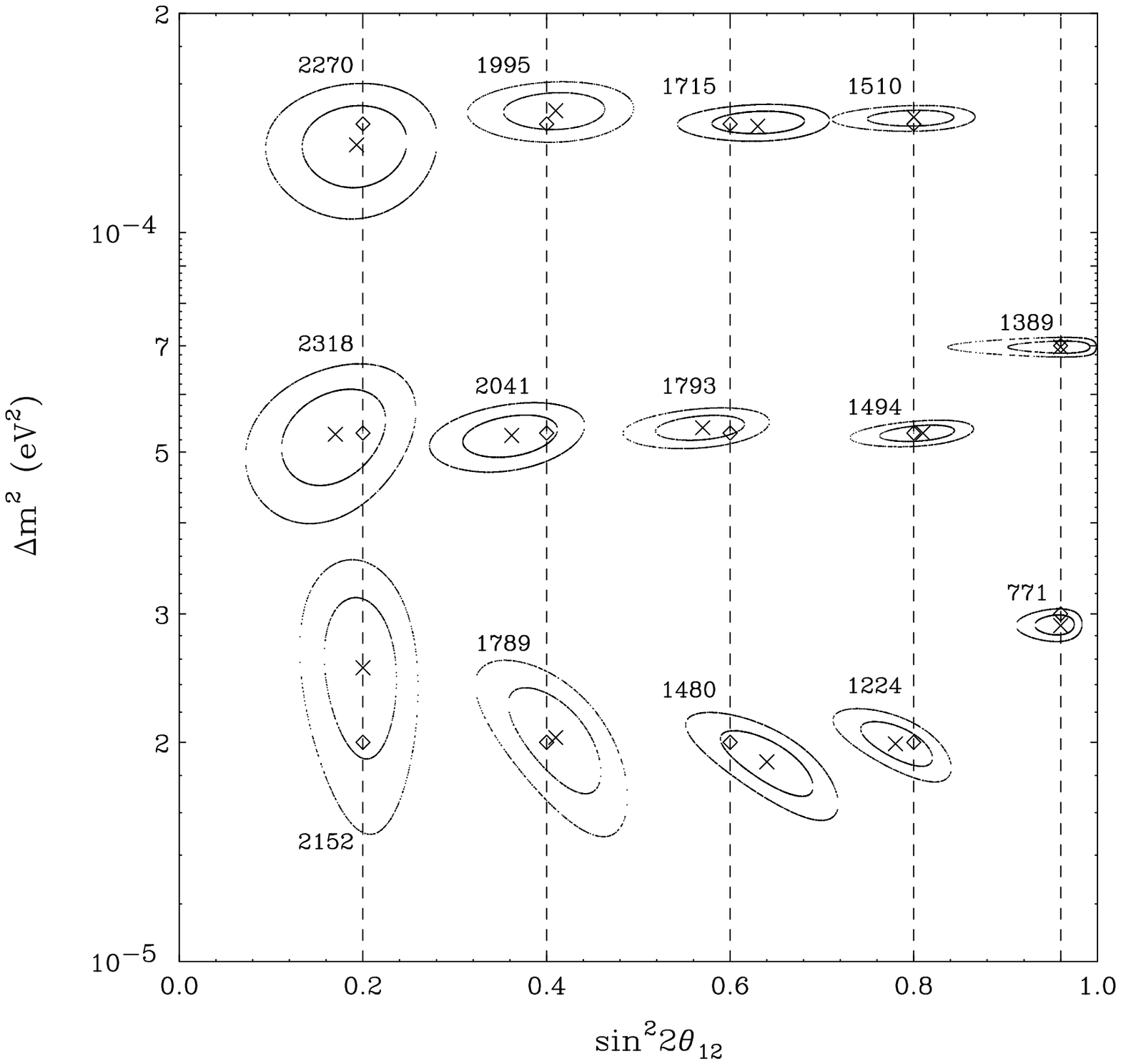,width=15cm,height=15cm}}
\caption{Fits to $e^+$ spectra for values of $\Delta m^2$ and $\rm{sin}^2 2\,\theta_{12}$ covering 
the entire region of the LMA solution with $\theta_{13}=0$. The 
$1\sigma$ (68.3\%) and $2\sigma$ (95.4\%) confidence contours are shown. The diamond is the
theoretical value for which data was simulated and the cross marks the best fit point. Each point
is labelled by the expected number of signal events in three years. If no oscillations occur, the
expectation is 2400 events. 
 }
\label{chi0}
\end{figure}

Figure~\ref{chi0} shows fits to $e^+$ spectra (expected after three years of running) 
for values of $\Delta m^2$ and $\rm{sin}^2 2\,\theta_{12}$ 
covering the entire region of the LMA solution with $\theta_{13}=0$.
 The plot shows $1\sigma$ (68.3\%) and $2\sigma$ (95.4\%) confidence contours. The diamond is the
theoretical value for which data was simulated and the cross marks the best fit point.
 Each point is labelled by the expected number of signal events.
The $\chi^2$ value for the best fit corresponds to a $\chi^2$ probability of at least 80\%.
For a given value of  $\Delta m^2$, the confidence regions get flatter
along the $\Delta m^2$-direction as $\rm{sin}^2 2\,\theta_{12}$ is increased from 0.2 to $1$, resulting
in a finer determination of $\Delta m^2$ for larger $\rm{sin}^2 2\,\theta_{12}$.
 The value of  $\rm{sin}^2 2\,\theta_{12}$
will be determined to an accuracy of $\pm 0.1$ ($2\sigma$) throughout almost all of the parameter space
under consideration.
 The $\Delta m^2$ value will be determined within $\pm 10$\%  ($2\sigma$) of its actual value for
 $10^{-5} \lsim \Delta m^2 \lsim 2 \times 10^{-4}\, \rm{eV}^2$ and \mbox{$\rm{sin}^2 2\,\theta_{12}>0.7$.}  

\begin{figure}[tb]
\mbox{\psfig{file=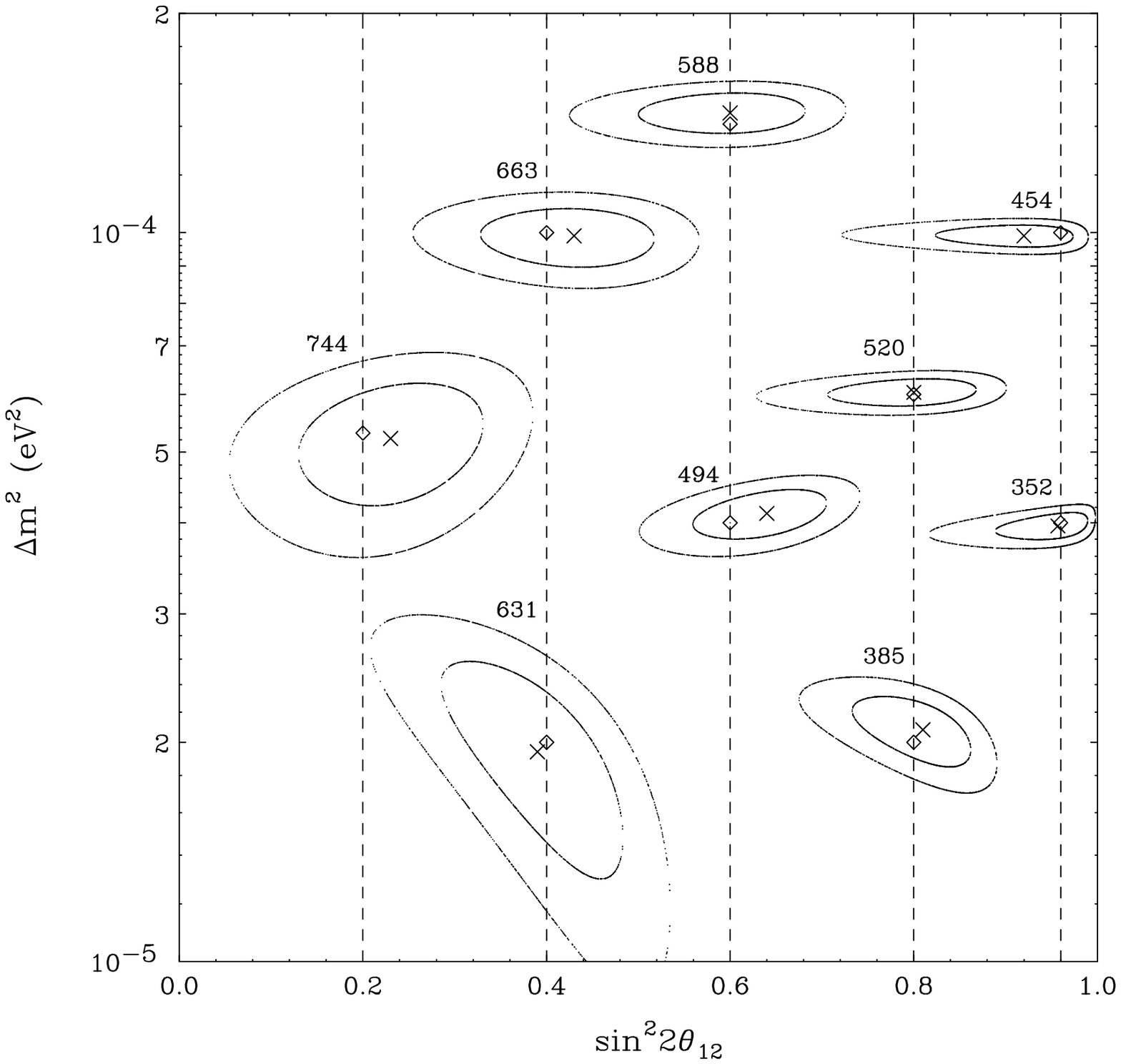,width=15cm,height=15cm}}
\caption{The same as Fig.~\ref{chi0} except that only one year of accumulated data is assumed. 
Correspondingly,  each point
is labelled by the expected number of signal events in one year. If no oscillations occur, the
expectation is 800 events.}
\label{chi2}
\end{figure}

We found that for  $\rm{sin}^2 2\,\theta_{13}=0.1$ the contours are almost identical to that
of  Fig.~\ref{chi0}; thus, KamLAND will not be sensitive to $\theta_{13}$ in any region
of the parameter space. In Fig.~\ref{chi2}
we show the expectation with one year of accumulated data (taking $\theta_{13}=0$) and find that
 in this time period KamLAND should already give us the LMA parameters to reasonably good accuracy. 
\section{Conclusion}
The results of our work are succinctly summarized in Figs.~\ref{chi0}--\ref{chi2}. Within
one year of operation, KamLAND measurements of the reactor $\bar{\nu}_e$ flux will establish whether 
or not the LMA solution (\mbox{$10^{-5} \lsim \Delta m_{21}^2 \lsim 10^{-4}\, \rm{eV}^2$} and 
$\rm{sin}^2 2\,\theta_{12}>0.3$) is correct.
If it is, in a span of three years
the accuracy with which $\rm{sin}^2 2\,\theta_{12}$, and especially
 $\Delta m_{21}^2$, will be determined is striking. KamLAND will give us
$\rm{sin}^2 2\,\theta_{12}$ to an accuracy of $\pm 0.1$ and $\Delta m_{21}^2$ to within a factor of
2 for $\rm{sin}^2 2\,\theta_{12}>0.2$; $\Delta m_{21}^2$ will be known to an accuracy of  
$\pm 10$\% for  $\rm{sin}^2 2\,\theta_{12}>0.7$  (at the $2\sigma$ level).
However, no knowledge
 will be gleaned about $\theta_{13}$. Our conclusions
 assume that the experiment employs a 1000 ton fiducial volume.
If the fiducial volume is instead 600 tons, 
 Fig.~\ref{chi2} applies for data collected over 20 months instead of 12 and 
the contours of Fig.~\ref{chi0} will be roughly twice as large
at the end of three years.  
Details of systematic
uncertainties and the accuracy with which the background shape is known will cause
minor changes in our results, but the overall conclusions will remain intact.  

If the LMA solution is found to be correct, the 
precision with which  $\Delta m_{21}^2$ will be pinned down at KamLAND will prove to be very beneficial
 to studies of $CP$-violation in the lepton sector in long baseline experiments~\cite{morecp,lfv}.

\vspace{0.25in}
\acknowledgements 
We thank A. Baldini, J. Learned and K. Whisnant for discussions.
This work was supported in part by a DOE grant No. DE-FG02-95ER40896 
and in part by the Wisconsin Alumni Research Foundation.
\vspace{0.25in}


\end{document}